\documentclass[8.5pt,twoside,twocolumn]{article}
\usepackage[super,sort&compress,comma]{natbib} 
\usepackage{mhchem}
\usepackage{times}
\usepackage{amsmath, amssymb, mathptmx} 
\usepackage{sectsty}
\usepackage{balance} 
\usepackage{graphicx,epsfig,array} 
\usepackage{lastpage}
\usepackage[format=plain,justification=raggedright,singlelinecheck=false,font=small,labelfont=bf,labelsep=space]{caption} 
\usepackage{fancyhdr}
\begin{document}

%
%
\newcommand{\etal}{\textit{et al. }}
\newcommand{\p}{\partial}
\renewcommand{\vec}[1]{\mathbf{#1}}
\newcommand{\taub}{\mbox{\boldmath$\tau$}}
\newcommand{\chib}{\mbox{\boldmath$\chi$}}
\newcommand{\xib}{\mbox{\boldmath$\xi$}}
\newcommand{\abs}[1]{\left|#1\right|}
\newcommand{\ept}[1]{\left\langle#1\right\rangle}
\newcommand{\trans}[1]{{#1}^{T}}
\newcommand{\unknown}{{??}}
\newcommand{\eg}{\textit{eg.} }
\newcommand{\deriv}[2]{\frac{\partial #1}{\partial #2}}
\newcommand{\ddt}[1]{\frac{\partial #1}{\partial t}}
\newcommand{\grad}[1]{\frac{\partial #1}{\partial \mathbf{x}}}
\newcommand{\floor}[1]{\lfloor #1 \rfloor}
\newcommand{\ceil}[1]{\lceil #1 \rceil}
\newcommand{\vbar}{\mathbf{\bar{v}}}
\newcommand{\xibar}{\mbox{\boldmath$\bar{\xi}$}}
\newcommand{\del}{\vec{\nabla}}
\newcommand{\thetab}{\mbox{\boldmath$\theta$}}
\newcommand{\sigb}{\mbox{\boldmath$\sigma$}}

\thispagestyle{plain}
\fancypagestyle{plain}{
\fancyhead[L]{\includegraphics[height=8pt]{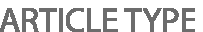}}
\fancyhead[C]{\hspace{-1cm}\includegraphics[height=20pt]{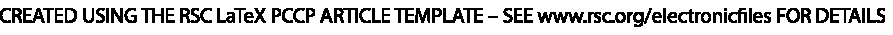}}
\fancyhead[R]{\includegraphics[height=10pt]{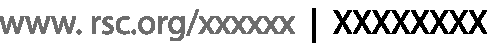}\vspace{-0.2cm}}
\renewcommand{\headrulewidth}{1pt}}
\renewcommand{\thefootnote}{\fnsymbol{footnote}}
\renewcommand\footnoterule{\vspace*{1pt}%
\hrule width 3.4in height 0.4pt \vspace*{5pt}} 
\setcounter{secnumdepth}{5}

\makeatletter 
\def\subsubsection{\@startsection{subsubsection}{3}{10pt}{-1.25ex plus -1ex minus -.1ex}{0ex plus 0ex}{\normalsize\bf}} 
\def\paragraph{\@startsection{paragraph}{4}{10pt}{-1.25ex plus -1ex minus -.1ex}{0ex plus 0ex}{\normalsize\textit}} 
\renewcommand\@biblabel[1]{#1}            
\renewcommand\@makefntext[1]%
{\noindent\makebox[0pt][r]{\@thefnmark\,}#1}
\makeatother 
\renewcommand{\figurename}{\small{Fig.}~}
\sectionfont{\large}
\subsectionfont{\normalsize} 

\fancyfoot{}
\fancyfoot[LO,RE]{\vspace{-7pt}\includegraphics[height=9pt]{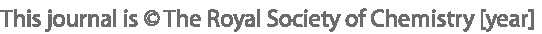}}
\fancyfoot[CO]{\vspace{-7.2pt}\hspace{12.2cm}\includegraphics{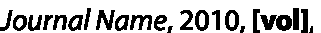}}
\fancyfoot[CE]{\vspace{-7.5pt}\hspace{-13.5cm}\includegraphics{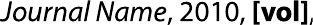}}
\fancyfoot[RO]{\footnotesize{\sffamily{1--\pageref{LastPage} ~\textbar  \hspace{2pt}\thepage}}}
\fancyfoot[LE]{\footnotesize{\sffamily{\thepage~\textbar\hspace{3.45cm} 1--\pageref{LastPage}}}}
\fancyhead{}
\renewcommand{\headrulewidth}{1pt} 
\renewcommand{\footrulewidth}{1pt}
\setlength{\arrayrulewidth}{1pt}
\setlength{\columnsep}{6.5mm}
\setlength\bibsep{1pt}

\twocolumn[
  \begin{@twocolumnfalse}
\noindent\LARGE{\textbf{Modeling the nonlocal behavior of granular flows down inclines}}
\vspace{0.6cm}

\noindent\large{\textbf{Ken Kamrin$^{\ast}$\textit{$^{a}$} and David L. Henann\textit{$^{b}$}}}\vspace{0.5cm}

\noindent\textit{\small{\textbf{Received Xth XXXXXXXXXX 20XX, Accepted Xth XXXXXXXXX 20XX\newline
First published on the web Xth XXXXXXXXXX 20XX}}}

\noindent \textbf{\small{DOI: 10.1039/b000000x}}
\vspace{0.6cm}

\noindent \normalsize{Flows of granular media down a rough inclined plane demonstrate a number of nonlocal phenomena. We apply the recently proposed nonlocal granular fluidity model to this geometry and find that the model captures many of these effects.  Utilizing the model's dynamical form, we obtain a formula for the critical stopping height of a layer of grains on an inclined surface.  Using an existing parameter calibration for glass beads, the theoretical result compares quantitatively to existing experimental data for glass beads.  This provides a stringent test of the model, whose previous validations focused on driven steady-flow problems.   For layers thicker than the stopping height, the theoretical flow profiles display a thickness-dependent shape whose features are in agreement with previous discrete particle simulations.  We also address the issue of the Froude number of the flows, which has been shown experimentally to collapse as a function of the ratio of layer thickness to stopping height.  While the collapse is not obvious, two explanations emerge leading to a revisiting of the history of inertial rheology, which the nonlocal model references for its homogeneous flow response.}
\vspace{0.5cm}
 \end{@twocolumnfalse}
  ]

\section{Introduction}


\footnotetext{\textit{$^{a}$~Department of Mechanical Engineering, MIT, Cambridge, MA, USA. E-mail: kkamrin@mit.edu}}
\footnotetext{\textit{$^{b}$~School of Engineering, Brown University, Providence, RI, USA. E-mail: david\_henann@brown.edu}}



Nonlocal effects in granular media manifest in myriad different ways.  At the origin of the nonlocality is the finite size of the grains themselves, inducing cooperative behaviors that defy local rheological description.  Examples include grain-size-dependent shear features in the steady flow profiles of granular media.  A local rheology can be extracted from uniform simple shearing data of a granular media \cite{dacruz05}; however, nonuniform steady flows  of the same material can be seen to violate such a relation \cite{jop08,koval09,kamrin10}, as the grain-size sets up an internal length-scale that effectively penalizes variations in flow-rate over space.  A more recently observed nonlocal manifestation is the mechanically-induced creep effect, also known as ``secondary rheology'' \cite{nichol10,reddy11,wandersman14}.  A highlight of this phenomenon is that flow anywhere in a granular media removes the yield condition everywhere, and the rheology of probes in the body varies with distance from the source of motion.  Some of the most compelling demonstrations of nonlocality in granular media can be observed in the behavior of grains on an inclined surface \cite{pouliquen99,silbert03,forterre03,midi04}.  Contrary to local rheological models, which predict a thickness-independent repose angle, experiments and discrete simulations verify that granular layers have a critical ``stopping height'' proportional to the grain size --- layers thinner than this value come to a stop, whereas thicker layers admit steady flow down the incline.

Recently, a nonlocal rheological model based on the concept of ``granular fluidity'' has shown itself able to reconcile the issues of grain-size dependent shear features as well as the mechanically-induced creep effect in granular media \cite{kamrin12,henann13,kamrin14,henann14,henann14b}.  While these demonstrations pertain primarily to the way in which grain size influences spatial fields in materials that are driven to flow, nonlocality as studied in inclined plane flows is of a relatively different nature, particularly the notion of a stopping height, i.e., whether a material flows at all.  Despite this distinction, in this paper we shall show that the nonlocal fluidity model captures the phenomenology observed in the inclined plane geometry.   Doing so provides a new test of the framework, in a family of problems distinct from those previously studied.

Herein, we compare theoretical results to known results  (i.e., experimental or DEM) on the issues of (i) the presence of a stopping height and how it varies with inclination, (ii) the variation in shape of the flow profile as inclination increases above the stopping height, and (iii) the dependence of  mean flow speed on the ratio of the thickness to the stopping height.  These three phenomena have been well-studied (see aggregated data in \citet{midi04}) and constitute the major manifestations of nonlocality in inclined plane flows.

\section{From local to nonlocal granular rheology}
We begin by describing the inertial granular rheology and the nonlocal granular fluidity model (per the review in \citet{kamrin14}).  The inertial rheology for steady flow of dense granular media relates the local stress state and the strain-rate \cite{dacruz05,jop05,jop06} and is expressed via the dimensionless relationship:
\begin{equation}\label{local_eqs}
\mu=\mu_{\rm loc}(I), \ \ \ I=\dot{\gamma}\sqrt{\rho_{\rm s}d^2/P}, \ \ \ \mu=\tau/P.
\end{equation}
In the above, $\mu$ is the ratio of shear stress $\tau$ and normal pressure $P$, and $I$ is the inertial number, where $\dot{\gamma}$ is the shear rate, $d$ is the mean grain diameter, and $\rho_{\rm s}$ is the grain density.  The function for $\mu_{\rm loc}(I)$ is empirically fit and is typically characterized by a yield coefficient $\mu_{\rm s}$ such that $\mu_{\rm loc}(I\to0)=\mu_{\rm s}$.  Under quasi-static circumstances, for which $\mu$ does not increase substantially above $\mu_{\rm s}$, a linear fit is often used,
\begin{equation}\label{linear}
\mu_{\rm loc}(I) = \mu_{\rm s} + b I,
\end{equation}
where $b$ is a dimensionless material parameter. As $\mu$ increases further above $\mu_{\rm s}$, curvature of the function is observed with an asymptote at an upper-limiting value of $\mu$, denoted $\mu_2$. A common fit is\cite{jop05}
\begin{equation}\label{nonlinear}
\mu_{\rm loc}(I) = \mu_{\rm s} + \Delta\mu/(I_0/I +1)
\end{equation}
where $\Delta\mu\equiv\mu_2-\mu_{\rm s}$ and $I_0\equiv  \Delta\mu/b$.

The inertial rheology works well in describing uniform flows (e.g., planar shearing) over a wide range of flow rates \cite{dacruz05}.  However, in non-uniform flow geometries in zones of slowly-flowing, quasi-static material the one-to-one inertial relation between $\mu$ and $I$ is violated \cite{koval09}.  The behavior displayed in these zones is definitively nonlocal; the bulk stress/strain-rate behavior at steady-state changes as the macroscopic geometry varies, even when the local kinematics are identical. 

The ``nonlocal granular fluidity'' (NGF) model may provide the solution to the above issues. It has demonstrated the ability to quantitatively predict granular flows in many disparate geometries, with predictions verified in 2D and 3D as compared to both discrete-particle simulations \cite{kamrin12} and experiments \cite{henann13}.  It is the first continuum model to quantitatively describe all experimentally-obtained flow data in the complex split-bottom family of geometries \cite{henann13}. It has also been shown to correctly capture other nonlocal phenomena such as nonlocally induced material weakening, whereby the motion of a boundary removes the yield strength of the material everywhere, permitting far-away loaded objects to creep through the grains when otherwise they would remain static \cite{henann14b}.

Our exisiting work has implemented the NGF equations in a reduced, steady-state-only form
\begin{equation}\label{NGF}
\begin{split}
&{\dot{\gamma}}=g \mu, \\[4pt]
&g_{\rm loc}(\mu,P)=\dfrac{\dot{\gamma}_{\rm loc}(\mu,P)}{\mu}=\dfrac{\mu_{\rm loc}^{-1}(\mu)\ \sqrt{P/\rho_{\rm s} d^2}}{\mu}, \\[4pt]
&g=g_{\rm loc}+\xi^2\nabla^2 g,
\end{split}
\end{equation}
where the linearized version of $\mu_{\rm loc}(I)$, Eq.\,\eqref{linear}, was used since flows of interest were all close to quasi-static. The field $g$ is a state parameter called the \emph{granular fluidity}, and $\xi$ is the plastic \emph{cooperativity length}, which is proportional to $d$.  Note that in planar shear, flow gradients vanish and the above reduces appropriately to the local rheology, but in the presence of gradients, the Laplacian term ``spreads'' fluidity based on $\xi$. In our previous work \cite{kamrin12,henann13}, we have verified that $\xi$ is, in fact, not a constant length but satisfies
\begin{equation}\label{ximu}
\xi(\mu) =\dfrac{A}{\sqrt{|\mu-\mu_{\rm s}|}}d ,
\end{equation}
roughly in-line with prediction of the kinetic-elasto-plastic (KEP) theory on which other fluidity models are based \cite{bocquet09}. The parameter $A$, the dimensionless \emph{nonlocal amplitude}, is the only new parameter in the model, which quantifies the spatial extent of cooperativity in the flow. 

\subsection{Dynamical system for fluidity}
In \citet{henann14}, we describe how the steady-state-only NGF model -- i.e.,  Eqs.\,\eqref{local_eqs}, \eqref{linear}, \eqref{NGF}, \eqref{ximu} -- is obtained in its entirety from the steady flow limit of a thermomechanically derivable dynamical system for $g$.  One treats the fluidity and its gradient as kinematic state variables with separate free-energy contributions.   By selecting the corresponding free-energies in a fashion that preserves the linear inertial rheology in uniform flow conditions, Eq.\,\eqref{linear}, we obtain the following system:
\begin{equation}\label{smallmunonlocal}
t_0 \dot{g}=A^2d^2\nabla^2 g - \left(\mu_{\rm s}-\mu\right)g - b\sqrt{\dfrac{\rho_{\rm s}d^2}{P}}\mu g^2,
\end{equation}
where $t_0>0$ is a constant time-scale.  The steady-state-only model arises as the approximate solution of the stable, steady behavior of $g$ in the above dynamical partial differential equation (PDE).  The approximation is valid as long as $g=g_{\rm loc}+\delta$ for some small function $\delta$, where $g_{\rm loc}$ emerges as the stable solution when the Laplacian term above vanishes.  To switch the local response to the more robust form, Eq.\,\eqref{nonlinear}, the same analysis can be reapplied giving
\begin{equation}\label{bigmunonlocal}
t_0 \dot{g}= A^2d^2\nabla^2 g - \Delta\mu \left(\dfrac{\mu_{\rm s}-\mu}{\mu_2-\mu}\right)g -b \sqrt{\dfrac{\rho_{\rm s}d^2}{P}}\mu g^2.
\end{equation}
When the above is reduced to a steady-state only model, the nonlocal system obtained matches the aforementioned one in the appropriate limit of $\mu$ near $\mu_{\rm s}$, as it should, with $\xi$ maintaining the same inverse square-root divergence behavior\footnote{The precise form of $\xi$ in the steady-state system corresponding to Eq.\,\eqref{bigmunonlocal} is
\begin{equation}\label{xinew}
\xi(\mu) = A\sqrt{\dfrac{\mu_2-\mu}{\Delta \mu|\mu-\mu_{\rm s}|}}d.
\end{equation}
Note that the limit of Eq.\,\eqref{xinew} as $\mu_2$ goes to infinity corresponds to Eq.\,\eqref{ximu}.}.

The dynamics of $g$ presented above have yet to account for bistability of granular flows nor do they account for the totality of transient effects that occur in a sheared granular media.  To the former point, recent experimental evidence suggests non-monotonicity of the local response  (contrary to Eqs \ref{linear} and \ref{nonlinear}) may exist giving rise to bistable flow hysteresis in certain circumstances \cite{dijksman11}.  To the latter point, Reynolds dilation during shear initiation induces transient variations in material strength, which are commonly described using critical-state models \cite{wroth} but which are not yet included in our fluidity dynamics.  Instead, our model above intends to provide is an accurate long-term dynamical behavior of the material when passing through a sequence of developed flowing states.  We note that the process of obtaining a steady-state-only relation from a fully dynamical form has a history within fluidity-based modeling  for other amorphous media \cite{bocquet09}.  The dynamical form of the nonlocal fluidity model shows mathematical similarities to order-parameter-based rheological approaches, which also feature a diffusing state variable \cite{aranson02}.

\section{Strengthening due to thinness}

\begin{figure}[!t]
\centering
\includegraphics{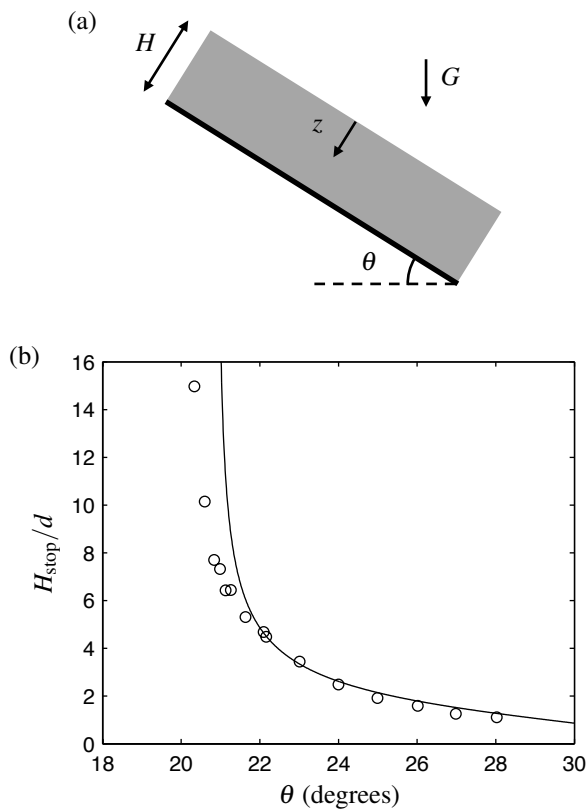}
\caption{(a) Setup of the inclined plane flow geometry. (b) Theoretically predicted $H_{\rm stop}$ locus (--), Eq.\,\eqref{hstoptheory}, as calibrated for glass beads on a fully rough base, compared to experimentally determined values ($\circ$) from \citet{pouliquen99}.}\label{hstopfig}
\end{figure}

In the inclined plane geometry, a layer of thickness $H$ is inclined to an angle $\theta$.  Assuming a packing fraction of $\phi$ and acceleration of gravity $G$, the stress distribution in the layer obeys
\begin{equation}
\mu=\tan\theta, \ \ \ \ P=\phi\rho_{\rm s}Gz\cos\theta.
\end{equation}
See Fig.\,\ref{hstopfig}(a) for reference.  Accordingly, if a local relation were applied, either Eq.\,\eqref{linear} or \eqref{nonlinear},  one would predict a universal angle of repose $\theta_{\rm r}=\arctan \mu_{\rm s}$, i.e., any layer tilted above $\theta_{\rm r}$ would be predicted to flow.

A signature of the cooperativity of granular motion is the fact that this universal repose angle is contradicted in experiments.  As shown initially by \citet{pouliquen99} and verified by others \cite{silbert03,forterre03,midi04}, the angle at which an initially flowing layer of grains comes to a stop, $\theta_{\rm stop}$, depends sensitively on the thickness of the layer when $H$ is small.  Inverting, one can extract a function $H_{\rm stop}(\theta)$ for every granular media and substrate, which represents the critical thickness at which a flowing layer at a certain angle would arrest.

Unlike our past studies, which have focused on nonlocal effects in media that is necessarily flowing, the problem of size-dependent strengthening of thin layers often presents conditions that do not satisfy the $g=g_{\rm loc}+\delta$ approximation necessary to reduce the mathematics to steady-state-only form; i.e., material stops ($g=0$) although the inclination can be steep enough for $g_{\rm loc}$ to be notably greater than zero.  To study inclined flows, we must revert to the primitive, dynamical form of the nonlocal fluidity model,  Eq.\,\eqref{bigmunonlocal}, to ensure accuracy of solutions, which should be valid up to and including steep inclination angles. 

We compare the results of our model to the experiments of \citet{pouliquen99}, where glass beads were used as the media.  The local continuum parameters for glass beads have been calibrated in \citet{jop05} and are $\mu_{\rm s}=0.382=\tan(20.9^{\circ})$, $\mu_2=0.643=\tan(32.76^{\circ})$, $b=\Delta\mu/I_0=0.938$, and $\rho_{\rm s}=2450\,$kg/m$^3$.  The nonlocal amplitude for glass beads was calibrated in \citet{henann13} to be $A=0.48$.  We can therefore apply our model in this geometry without additional parameter fitting, as long as we can identify accurate boundary conditions for the fluidity and basal slip velocity.    We have tried two different fluidity boundary conditions in our past studies, chosen primarily based on simplicity, with the understanding that the choice is less important far from system boundaries due to the source-diffusion nature of the fluidity system.  However, due to the thinness of the layers we wish to consider here, the accuracy of the boundary condition has a much larger influence on the flow behavior.  To remove this issue, we opt instead to extract the fluidity boundary conditions directly from existing flow data. In discrete particle simulations of \citet{silbert03}, who also used spherical particles, it is apparent that adjacent to a fully rough boundary, the shear-rate (and velocity fluctuations) approach an approximately vanishing state, and the mean velocity vanishes, indicating no slip.  At or near the free surface, there is usually (though not always) a zone where the shear-rate appears to level off and a vanishing shear-rate gradient is observed.  Likewise, we will assume $g=0$ and $v=0$ along the rigid surface at $z=H$ and take $\partial g/\partial z = 0$ at the free surface, $z=0$.  Since these boundary conditions are homogeneous, they always permit both flowing and static solutions.

The phase diagram of flowing and non-flowing states according to the nonlocal model can be obtained by determining the conditions necessary for the global $g=0$ solution to be stable, as this decides if a system perturbed to flow returns to a static state.  In Eq.\,\eqref{bigmunonlocal}, a perturbation about $g=0$ renders the $g^2$ term negligible.  The remaining PDE is linear and, substituting $\mu=\tan\theta$, it is solved by
\begin{multline}
g(z,t) = \\[4pt] C\exp\left[{\left(-\dfrac{\pi^2A^2d^2}{4 H^2} + \Delta \mu\dfrac{\tan\theta-\mu_{\rm s}}{\mu_2-\tan\theta} \right) \dfrac{t}{t_0}} \right] \cos\left(\dfrac{\pi z}{2H}\right),
\end{multline}
where $C$ is a constant prefactor. The solution decays to $g=0$ as long as the term in the exponent is negative.  Hence, a perturbed layer solidifies if
\begin{equation}\label{hstoptheory}
H< \dfrac{\pi Ad}{2}  \sqrt{\dfrac{\mu_2-\tan\theta}{\Delta\mu(\tan\theta-\mu_{\rm s})}}=H_{\rm stop}(\theta).
\end{equation}
A similar mathematical technique was utilized in \cite{aranson02}. The above shows that for tall layers, $H\gg O(d)$, the material is predicted to have a common repose angle of $\theta_1=\arctan(\mu_{\rm s})$.  However, for thinner layers, an inclination higher than this value is needed to stop flow.  This applies up to an upper limit of $\theta_2=\arctan(\mu_2)$ at which all layers, independent of height, are predicted to flow.  The existence of two critical angles having these properties has been verified as a common trait of $H_{\rm stop}$ data in inclined plane flows in multiple experiments and particle simulations involving different materials and substrates \cite{midi04}.  When it comes to specific functional forms for $H_{\rm stop}$ data, several fit curves have been proposed, a topic we shall return to later, but all invoke critical $\theta_1$ and $\theta_2$ values as described here. It bears noting that the derived $H_{\rm stop}$ relation, Eq.\,\eqref{hstoptheory}, is linearly proportional to the cooperativity length, Eq.\,\eqref{xinew}, i.e., $H_{\rm stop}(\arctan\mu)=(\pi/2)\xi(\mu)$. This is in line with experimental observations of \citet{pouliquen04}, who reported that velocity correlation lengths in inclined plane flows vary with $\theta$ (and hence $\mu$) in a similar manner as $H_{\rm stop}$. We emphasize that this is a derivable consequence -- not an underlying assumption -- of the dynamical system, Eq.\,\eqref{bigmunonlocal}. Further, one will obtain the same result when working with the simplified dynamical system, Eq.\,\eqref{smallmunonlocal}, albeit with slightly modified functional forms.

For a direct comparison, in Fig.\,\ref{hstopfig}(b), the above predicted form for $H_{\rm stop}(\theta)$, using the parameters for glass beads,  is compared against Pouliquen's experimentally determined values using glass beads with a fully rough base.  It is worth pointing out that the model's result was obtained using the same continuum parameters that were used to successfully predict steady flow fields of glass beads  in split-bottom cells and other geometries \cite{henann13}. Yet here, the question is of a different nature, one of predicting input conditions for flow stoppage rather than velocity profiles in a flowing body.

We note that a size-dependent starting height $H_{\rm start}(\theta)$ -- the height at which a static layer initiates flow --  is also observed experimentally.  Its curve depends on the preparation of the static layer and differs somewhat from $H_{\rm stop}$, though it shares the same qualitative features (e.g., $\theta_1$ and $\theta_2$ values where the curve asymptotes and vanishes respectively).  Our analysis above is tailored to $H_{\rm stop}$ since it reflects the limiting behavior of a flowing solutions as $H$ (or $\theta$) is decreased, however we note that without a bistable term in our dynamics, a distinct $H_{\rm start}$ curve does not arise theoretically.

\section{Velocity profiles}
\begin{figure}[!t]
\centering
\includegraphics{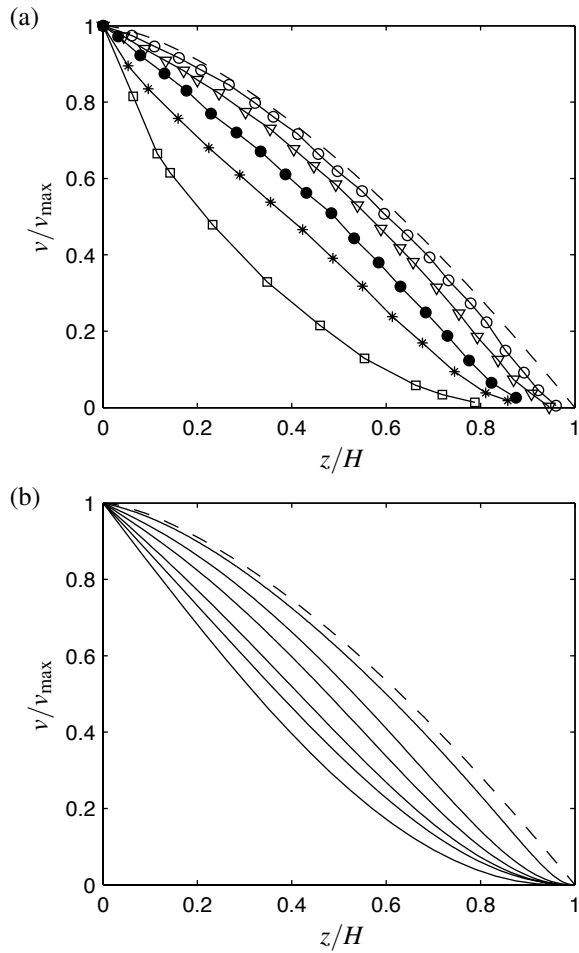}
\caption{(a) Discrete particle simulation data of \citet{silbert03} for $\theta=24^{\circ}$ and many layer thicknesses (increasing from top to bottom).  Dashed line is the Bagnold profile. (b) Theoretical profiles for the same relative heights, $H/H_{\rm stop}\cong$ 1.1, 1.9, 2.7, 4.1, 6.4, as well as  $H/H_{\rm stop}=20$.}\label{profiles}
\end{figure}

For angles between $\theta_1$ and $\theta_2$, and $H$ exceeding $H_{\rm stop}$, the material has a steady flow state down the incline. As reported collectively in GDR \citet{midi04} and explored directly in \citet{silbert03}, the shape of the flow profile varies from concave up, to roughly linear, to concave down as $H$ is increased.  The concave down shape observed for large $H$ is well-fit by the so-called "Bagnold profile" obtainable by integrating the strain-rate predicted by the inertial rheology through the thickness of the layer.  The result is of the form
\begin{equation}\label{Bagnold}
v_{\rm Bag}(z)\propto H^{3/2}-z^{3/2}.
\end{equation}
Figure \ref{profiles}(a) shows the discrete particle simulation data of Silbert \cite{silbert03} for inclination angle $\theta=24^{\circ}$ and various values of $H>H_{\rm stop}(\theta=24^{\circ})$.  We have computed numerical solutions to our model at the same inclination angle and several layer thicknesses.  We note that the stopping curves of the simulated material and glass beads are different;  at $\theta=24^{\circ}$, the simulated particles have an $H_{\rm stop}/d$ value more than twice that of the experiment.  To make an appropriate albeit qualitative comparison, we select our $H$ values so as to (roughly) match the relative heights $H/H_{\rm stop}$ used in the discrete simulations.  The results are in Fig.\,\ref{profiles}(b).  The model predictions are calculated by evolving Eq.\,\eqref{bigmunonlocal} using finite-differences in MATLAB, using $\Delta x \ll d$ and the Bagnold profile as the initial guess\footnote{In the numerics, we allow a small pressure on the top surface corresponding to the weight of a layer of $(1/2)d$ thickness, i.e., $P_{\rm top}=(1/2)\phi\rho_{\rm s}Gd\cos\theta$, to ensure that the coefficient of the $g^2$ term in Eq.\,\eqref{bigmunonlocal} remains bounded.}.

The model shows that for $H$ near $H_{\rm stop}$ the profiles display a concave-up feature in agreement with the particle simulations.  It could be said that the concave up appearance is due to the comparative unimportance of the $g^2$ term in Eq.\,\eqref{bigmunonlocal} when flows are slow, which, if neglected, gives a cosinusoidal solution for $g$ and $v\sim 1-\sin(z\pi/2H)$.  As $H$ increases to be much greater than $H_{\rm stop}$, the profiles approach the Bagnold profile, which agrees with the discrete simulations and is sensible since the relative importance of the particle-size-dependent term in Eq.\,\eqref{bigmunonlocal} diminishes in a tall flow; if neglected, the remaining terms give exactly the inertial rheology at steady flow.  The transition between these extremes is marked by velocity profiles displaying a linear character.  It can also be observed that, in agreement with the data and in contrast to the Bagnold profile, the model does not always require a zero strain-rate at the free surface.

\section{Discussion of the Froude number}

\begin{figure*}[!t]
\centering
\includegraphics{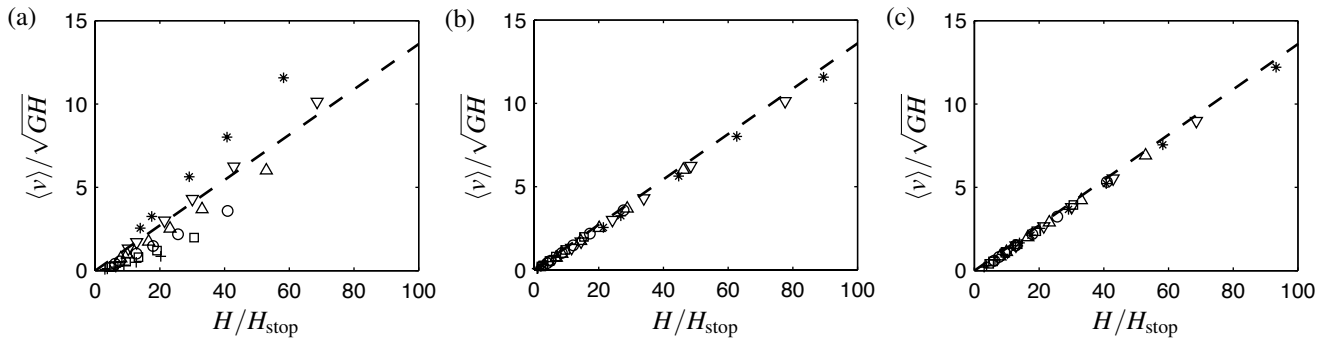}
\caption{Predicted Froude number, $\langle v\rangle/\sqrt{GH}$, at inclinations $\theta= 22.5 (+)$, $24 (\Box)$, $25.5  (\circ)$, $27 (\bigtriangleup)$, $28.5  (\bigtriangledown)$, and $30 (*)$ degrees while varying $H$ from $12$-$80d$, plotted against the relative height, utilizing $H_{\rm stop}(\theta)$ as shown in (a) Eq.\,\eqref{hstoptheory} and (b)  Eq.\,\eqref{hstopexp}. (c) Theoretical result when the local rheology is chosen consistent with the theoretical $H_{\rm stop}$, per Eq.\,\eqref{lochstop}. In (a)-(c), dashed line is the experimental collapse line, Eq.\,\eqref{collapse}, with $\beta = 0.136$.}\label{froude_fig}
\end{figure*}

We now discuss the mean flow speed of granular media down an incline. \citet{pouliquen99}, and others thereafter \cite{silbert03,forterre03,midi04}, have observed that the Froude number of the flow, $\langle v \rangle/\sqrt{GH}$, appears to be a relatively well-defined, one-to-one function of the relative height,  $H/H_{\rm stop}$, for all angles and heights as long as $H$ is not close to $H_{\rm stop}$.  In this range, they find
\begin{equation}\label{collapse}
\dfrac{\langle v \rangle}{\sqrt{GH}}\approx\beta \dfrac{H}{H_{\rm stop}},
\end{equation}
where, for glass beads, $\beta= 0.136$. When results of the nonlocal theory are plotted in this fashion, Fig.\,\ref{froude_fig}(a), we do not observe a one-to-one correspondence between the Froude number and the relative height.  We have two comments on this point, that may explain the discrepancy.

\subsection{A more precise $H_{\rm stop}$ curve}
Collapse of the Froude number is sensitively dependent on the fit one uses for $H_{\rm stop}(\theta)$.  In the form obtained from our theory, we notice some deviations from the experiment which magnify, unsurprisingly, as one approaches the asymptote at $\theta_1$ (see Fig.\,\ref{hstopfig}).  A different fit function\cite{forterre03}, empirically matched to the data in \citet{pouliquen99}, takes the form
\begin{equation}\label{hstopexp}
H_{\rm stop}(\theta)=L_0\left(\dfrac{\mu_2-\mu_{\rm s}}{\tan\theta-\mu_{\rm s}}-1\right),
\end{equation}
where $L_0/d\approx1.65$ (see the caption to Fig.\,8 of \citet{forterre03}). If we use the above relation for $H_{\rm stop}$, a rather strong collapse of the Froude number versus relative height emerges (see Fig.\,\ref{froude_fig}(b)).  Thus, it may be that the lack of a collapse of the Froude number is attributable to deviation of the predicted $H_{\rm stop}$ curve.

\subsection{Consistent local rheology}
An alternative explanation could be found if we revisit the history of the local inertial rheology.  The rheology in Eq.\,\eqref{nonlinear} was arrived at, coincidentally, by fitting data for glass beads flowing down inclines (see Appendix A of \citet{jop05}).  First, a form for $H_{\rm stop}$ was fitted empirically from experiments (i.e., Eq.\,\eqref{hstopexp}). Then $\langle v \rangle$ was calculated in terms of an as-yet-unknown function $\mu_{\rm loc}(I)$ (assuming fully local rheology and, consequently, the Bagnold velocity profile, Eq.\,\eqref{Bagnold}). This result was subsequently substituted, along with the empirical $H_{\rm stop}$ function, into Eq.\,\eqref{collapse} to solve for $\mu_{\rm loc}(I)$.  The result is a local rheology having the form
\begin{equation}\label{lochstop}
\begin{split}
I=\mu_{\rm loc}^{-1}(\mu)&=\dfrac{5\beta}{2\sqrt{\phi \cos(\arctan{\mu})}}\dfrac{d}{H_{\rm stop}(\arctan{\mu})}\\[4pt]
&\approx \dfrac{d}{B\ H_{\rm stop}(\arctan{\mu})},
\end{split}
\end{equation}
which generates Eq.\,\eqref{nonlinear}, upon substituting Eq.\,\eqref{hstopexp} for $H_{\rm stop}$.  As was done in \citet{jop05}, the term containing the square-root is replaced with a constant $B$ ($\approx 2.17$ for glass beads) because the $\cos$ function does not vary much in the range of use. 
We see that, historically, the fit function chosen for $H_{\rm stop}$ gives rise to the functional form one obtains for the inertial rheology.

Let us repeat this logic in a thought experiment.  Suppose, through Eq.\,\eqref{lochstop}, we replace the local rheology in our theory with the one corresponding to our fit for the $H_{\rm stop}$ function, Eq.\,\eqref{hstoptheory}.  The result is
\begin{equation}\label{localnew}
I=\mu_{\rm loc}^{-1}(\mu)=\dfrac{2\sqrt{\Delta \mu}}{\pi AB}\sqrt{\dfrac{\mu-\mu_{\rm s}}{\mu_2-\mu}}.
\end{equation}
With this, the nonlocal fluidity dynamics are expressible conveniently as
\begin{multline}\label{nonlocalnew}
t_0 \dot{g}= A^2d^2\nabla^2 g \\[4pt] +  \dfrac{\pi^2A^2}{4} \left[\left(\dfrac{d}{H_{\rm stop}}\right)^2 g - B  \sqrt{\dfrac{\rho_{\rm s}d^2}{P}}\left(\dfrac{d}{H_{\rm stop}}\right) \mu g^2\right].
\end{multline}
Only the $g^2$ term has changed so the previous stability argument identically produces our $H_{\rm stop}$ function, and Eq.\,\eqref{localnew} is now obtained in the homogeneous flow limit as the new local response\footnote{In fact, the steady-state-only system corresponding to Eq.\,\eqref{nonlocalnew} is given through Eq.\,\eqref{NGF} with local rheology Eq.\,\eqref{localnew} and cooperativity length Eq.\,\eqref{xinew}. We note that the coefficient of $g^2$ in Eq.\,\eqref{nonlocalnew} is for the moment only defined for $\mu>\mu_s$ as needed for the Froude number analysis, but any monotonic extension of $1/H_{\rm stop}$ for $\mu<\mu_{\rm s}$ may be assumed without affecting the analysis nor the corresponding steady-state-only relation.}.  

We apply the above system to inclined plane flows.  Letting $\tilde{z}=z/H$ and $\tilde{g}=g\mu H_{\rm stop}/\sqrt{GH}$, writing $P=\phi\rho_{\rm s}Gz\cos\theta$, recalling the definition of $B$, and allowing $\theta_1\le\theta\le\theta_2$, Eq.\,\eqref{nonlocalnew} produces steady solutions given by
\begin{equation}\label{newform}
0= \tilde{g}_{\tilde{z}\tilde{z}}  +  \dfrac{\pi^2}{4}\left(\dfrac{H}{H_{\rm stop}}\right)^2 \left( \tilde{g} - \dfrac{2}{5\beta} \sqrt{\dfrac{1}{\tilde{z}}}\tilde{g}^2  \right).
\end{equation}
The only varying parameter in the above is the ratio $H/H_{\rm stop}$.  Hence, all solutions have the form, $\tilde{g}=\tilde{g}(\tilde{z};H/H_{\rm stop})$.  Likewise, the velocity field obeys
\begin{multline}
v(z)=\int_z^H \dot{\gamma}(z') \ dz' = \int_{z}^H\mu g(z')\ dz' \\[4pt]=  \sqrt{GH}\ \tilde{V}(z/H;H/H_{\rm stop}).
\end{multline}
Finally, we arrive at a relation for the Froude number:
\begin{multline}
\dfrac{\langle v \rangle}{\sqrt{GH}}  = \dfrac{1}{\sqrt{GH}}\dfrac{1}{H}\int_{0}^H v(z')\ dz' \\[4pt]=  \int_{0}^1 \tilde{V}(\tilde{z}; H/H_{\rm stop})\ d\tilde{z}  = {\rm Fr}(H/H_{\rm stop}).
\end{multline}
We conclude that when the same $H_{\rm stop}$ curve we derive theoretically is used to fit the local inertial relation, the emergent nonlocal theory \emph{requires} a collapse between the Froude number and $H/H_{\rm stop}$ for all $H>H_{\rm stop}$.  This is an interesting result because while the local relation on its own would require such a collapse by definition, this is the full nonlocal solution.  When $H/H_{\rm stop}$ is large enough, the second-derivative term in Eq.\,\eqref{newform} is negligible (excepting a rapid variation near the lower boundary), and the solution for $\tilde{g}$ approaches that of a purely local material behavior.  Likewise the function ${\rm Fr}(H/H_{\rm stop})$ obeys ${\rm Fr}(H/H_{\rm stop})\sim \beta H/H_{\rm stop}$ for $H\gg H_{\rm stop}$, matching Eq.\,\eqref{collapse}.  Numerical integration of Eq.\,\eqref{nonlocalnew} verifies these analytical properties (see Fig.\,\ref{froude_fig}(c)).  We observe that ${\rm Fr}(1)=0$, which is sensible as this corresponds to the stopping height, and our model does not include bistable terms.  Although it is unclear in discrete simulations and experiments what the precise behavior of the Froude number is near $H/H_{\rm stop}=1$, for round grains on a fully rough surface many data trends show the Froude number exceeds zero at this height \cite{pouliquen99} and other data suggests it equals zero \cite{weinhart12}, as our model purports.

A short commentary is in order.  The above result comes at the cost of adjusting the $\mu_{\rm loc}$ relation from the commonly used form of Eq.\,\eqref{bigmunonlocal}, to a form compatible with the theory's own predicted $H_{\rm stop}$ function.  The new relation, Eq.\,\eqref{localnew}, is qualitatively similar to the previous, with $\mu$ increasing monotonically from $\mu_{\rm s}$ to $\mu_2$ as $I$ increases.  One difference is that the new relation has $d\mu_{\rm loc}/dI=0$ at $I=0$ whereas the former has a positive slope.  There has been a debate recently over the behavior of $\mu_{\rm loc}$ near $I=0$.  Some experimental data suggests, in fact, a negative initial slope for $\mu_{\rm loc}$ \cite{dijksman11, kuwano13}, while other fits suggest a steeper-than-linear power-law behavior near $I=0$ \cite{peyneau08}.  It bears noting that our previous work on nonlocal fluidity has focused solely on quasi-static flows, in which $\mu<\mu_{\rm s}$ almost everywhere and the only important aspect of the local rheology is the value of $\mu_{\rm s}$.


\balance

\section{Conclusion}
We have applied the theory of nonlocal granular fluidity to the canonical problem of size-dependence in granular inclined plane flows.  The theory, as calibrated to glass beads based on prior data, predicts an $H_{\rm stop}$ curve that matches experimental data rather well.  Further, the theory predicts flow profiles that vary in shape in a fashion consistent with existing discrete simulation data, marked by an upward curvature for layer thicknesses $H$ close to $H_{\rm stop}$ and transitioning to the Bagnold profile for large $H$.  While the Froude number does not exhibit a direct collapse against $H/H_{\rm stop}$ we have identified two possible explanations for this and corresponding ways in which the collapse can indeed be obtained.

Perhaps the most compelling result herein is the prediction of the $H_{\rm stop}$ function, which suggests the nonlocal fluidity concept could be used to model other size-sensitive flow stoppage phenomena, such as silo jamming.  The famous Beverloo correlation \cite{beverloo61}, an empirical functional form that gives silo flow rate in terms of aperture opening size, indicates a critical opening size at which flow stops.  This effect may be roughly analogous to the stopping height observed for inclined flows, and it would be useful to apply the nonlocal fluidity model in this geometry to determine if the Beverloo correlation can be obtained from the nonlocal model.

\section*{Acknowledgements}\label{ack}
KK acknowledges funds from NSF-CBET-1253228 and the MIT Department of Mechanical Engineering, and DLH acknowledges funds from the Brown University School of Engineering.


\footnotesize{

\providecommand*{\mcitethebibliography}{\thebibliography}
\csname @ifundefined\endcsname{endmcitethebibliography}
{\let\endmcitethebibliography\endthebibliography}{}

}

\end{document}